\documentclass[11pt]{article}

\usepackage[dvips]{graphicx}
\usepackage{xspace}
\graphicspath{{./figs/}}
\usepackage{moriond,epsfig}

\newcommand{\SK}{Super-Kamiokande\xspace}
\newcommand{\nue}{$\nu_{e}$\xspace}
\newcommand{\numu}{$\nu_{\mu}$\xspace}
\newcommand{\nutau}{$\nu_{\tau}$\xspace}
\newcommand{\nus}{$\nu_{s}$\xspace}
\newcommand{\dm}{\Delta m^2}
\newcommand{\SinSq}{\sin^2 2\theta}
\newcommand{\eV}{\mbox{eV}^2}
\newcommand{\Flux}{$\Phi_{\rm ND}$\xspace}
\newcommand{\FN}{$F/N$\xspace}
\newcommand{\eSK}{$\epsilon_{\rm SK}$\xspace}
\newcommand{\oneRmu}{1R$\mu$\xspace}
\newcommand{\oneRe}{1R$e$\xspace}
\newcommand{\Dqp}{\Delta\theta_p}
\newcommand{\Rnqe}{$R_{\rm nqe}$\xspace}
\newcommand{\Ltot}{{\cal L}_{\rm tot}}
\newcommand{\Lnorm}{{\cal L}_{\rm norm}}
\newcommand{\Lsys}{{\cal L}_{\rm sys}}
\newcommand{\Lshape}{{\cal L}_{\rm shape}}
\newcommand{\Nobs}{N_{\rm obs}}
\newcommand{\Nexp}{N_{\rm exp}}
\newcommand{\Enurec}{E_{\nu}^{\rm rec}}
\newcommand{\JHFnu}{JHF-$\nu$\xspace}
\begin{document}
\vspace*{4cm}
\title{THE RESULTS OF OSCILLATION ANALYSIS IN K2K EXPERIMENT\\
AND\\
AN OVERVIEW OF JHF-{\boldmath $\nu$} EXPERIMENT}
\author{Issei Kato\\
        for the K2K collaboration}
\address{Department of Physics, Faculty of Science,
         Kyoto University, Sakyo-ku, Kyoto 606-8502, Japan}
\maketitle
\abstracts{
This paper presents the results of oscillation analysis in K2K
experiment. The results show indications of neutrino oscillation and give
a new constraint on the oscillation parameters. The difference of neutrino
masses squared $\Delta m^2$ lies between 1.5 and
3.9$\times10^{-3}~\mbox{\rm eV}^2$ at $\sin^2 2\theta=1$ with the
confidence level of 90\%. In addition to these results, a brief overview of
future long-baseline neutrino experiment in Japan, JHF-$\nu$ experiment,
is also given in this paper.
}

\section{Introduction}
Since a discovery of neutrino oscillation in atmospheric neutrinos by
\SK (SK)~\cite{SK:evidence}, a lot of attention has been directed to this new
phenomena. The zenith angle distribution of atmospheric neutrinos
observed in SK shows a clear deficit of upward-going \numu, which is
well explained by two-flavor \numu--\nutau oscillation with
$\dm\sim3\times10^{-3}~\eV$ and $\SinSq\sim1$.

The KEK-to-Kamioka long-baseline neutrino experiment (K2K) uses an
accelerator-produced neutrinos with a neutrino flight length of 250~km
to probe the same $\dm$ region as that explored by atmospheric
neutrinos. Recently, we have performed analysis for neutrino oscillation
using the information of \numu flux at SK together with the shape of
energy spectrum. This analysis is based on the data taken from June,
1999 to July 2001 (K2K-I) corresponding to $4.8\times10^{19}$ protons on
target (POT), which is approximately half of allocated POT. Since there
already exist a lot of descriptions on K2K
experiment~\cite{K2K:detection,K2K:SciFi-NIM,K2K:MRD-NIM,K2K:other},
only the procedure of oscillation analysis and its results are presented
in this paper. The description in this paper is based on the
publication~\cite{K2K:oscillation}.

While the first generation neutrino oscillation experiments have 
successfully being done and the oscillations in the atmospheric and
solar region are confirmed by several experiments, \numu--\nue
oscillation has not been seen except for LSND and is still in mystery.
In order to discover the \numu--\nue oscillation and to open possibilities
for detailed studies of neutrino mass ,lepton sector mixing and farther
possibilities for discovery of new physics, JHF-to-Kamioka
long-baseline neutrino experiment, JHF-$\nu$, is planned in Japan as a
next generation neutrino experiment. A brief overview of JHF-$\nu$
experiment is given in the last section.

\section{Strategy for Oscillation Analysis in K2K}
The neutrino beam in K2K experiment is produced by a 12~GeV proton beam
from the KEK proton synchrotron. After protons hit an aluminum target,
the produced positively charged particle, mainly $\pi^+$, are focused
by a pair of pulsed magnetic horns~\cite{K2K:horn}. The pion momentum
and angular distributions downstream of second horn are occasionally
measured by gas-Cherenkov detector (PIMON)~\cite{K2K:maruD} in order
to verify the beam Monte Carlo (MC) simulation and to estimate the
errors on the flux prediction at SK~\cite{K2K:detection}. The neutrino
beam produced from the decays of these particles in a decay pipe of
200~m long are 98\% pure muon neutrinos with a mean energy of 1.3 GeV.
Near neutrino detectors (ND) are located at 300~m from the proton target.
ND measure the stability and direction of the neutrino beam as well as
neutrino flux and energy spectrum just after the production. SK is
located at 250~km from KEK. Here the neutrino flux and energy spectrum
after neutrinos travel the length of 250~km are measured.
The measurements at ND are extrapolated to SK by multiplying far-to-near
flux ratio (\FN). The detailed description for \FN can be found in the
reference~\cite{K2K:detection}.

In the case of the presence of neutrino oscillation, both the reduction
in the number of neutrino events and the distortion in the energy
spectrum are expected after neutrinos travel a fixed flight
length, because the oscillation probability depends on neutrino energy:
\begin{equation}
P(\nu_\mu\to\nu_x)=\SinSq\sin^2\frac{\dm L}{4E_\nu},
\end{equation}
where $\theta$ is mixing angle, $\dm$ is the difference of neutrino
mass squared, $E_\nu$ is neutrino energy, and $L$ is the neutrino
flight length which is fixed to be 250~km in the K2K case.
Therefore both the number of observed events and the spectral
shape information at SK are used to compare them with the expectation
from ND measurements. In this analysis, all the beam induced neutrino
events observed in the fiducial volume of SK are used to measure the
overall suppression of flux, while in order to enhance the fraction of
charged-current (CC) quasi-elastic (QE) interactions
($\nu_\mu+n\to\mu+p$), 1-ring $\mu$-like events (\oneRmu) are used to
study the spectral distortion. Since the proton momentum in the CCQE
events is typically below Cherenkov threshold, only muon is visible.
Assuming QE interaction in \oneRmu and neglecting Fermi momentum, the
neutrino energy can be reconstructed as 
\begin{equation}
 \Enurec
 =\frac{m_{N} E_{\mu} + m_{\mu}^{2}}{m_{N} - E_{\mu} + p_{\mu}
 \cos\theta_{\mu}},
\end{equation}
where $m_{N}$, $E_{\mu}$, $m_{\mu}$, $p_{\mu}$ and $\theta_{\mu}$ are
nucleon mass, muon energy, muon mass, muon momentum and muon scattering
angle with respect to the direction of neutrino beam, respectively.

\section{Measurements of Neutrino Flux and Spectrum at ND}
The ND consist of two detector systems: a 1~kilo-ton water Cherenkov
detector (1KT) and a fine-grained detector (FGD). The flux normalization
is measured by 1KT to estimate the expected number of events at
SK. Since 1KT has the same detector technology as SK, the most of the
systematic uncertainties on the measurements, mainly on the detection
efficiency and cross-section of interactions, are canceled in comparison
between ND and SK measurements. The energy spectrum is measured by
analyzing the muon momentum and angular distributions in the both
detector systems. The 1KT has high efficiency to reconstruct muons with
the momentum of below 1 GeV/$c$ and full $4\pi$ coverage in solid
angle. However, the 1KT has little efficiency to reconstruct muons with
the momentum of above 1.5 GeV/$c$ since they exit the detector. On the
other hand, the FGD has high efficiency to measure the muons above 1
GeV/$c$, the relevant energy range can be covered by these two
complementary detectors.

\subsection{1KT Analysis}
\label{sec:1KT}
In the 1KT analysis, the \numu interactions are measured by detecting
Cherenkov light emitted from produced charged particles. The vertices,
direction and momentum of Cherenkov rings are reconstructed with the
same method as in SK~\cite{SK:method}. The event selection criteria for
the measurement of integrated flux are the same as those in the
reference~\cite{K2K:detection}: (a) Deposited energy is greater
than 100~MeV. (b) The reconstructed vertex is inside the 25~t fiducial
volume which is defined by a 2~m radius, 2~m long cylinder oriented to
the beam axis, in the upstream side of the detector. The expected
number of events with the vertices fully contained (FC) in SK fiducial
volume is estimated to be $80.1^{+6.2}_{-5.4}$. The correlation
between energy bins from the spectrum measurement at ND and \FN are
taken into account in the estimation of the errors, described in detail
below. The major contribution to the errors come from the uncertainties
in the \FN (${}^{+4.9}_{-5.0}\%$) and the normalization ($\pm5\%$). The
normalization error is dominated by uncertainties of the fiducial
volumes due to the vertex reconstruction both in 1KT (4\%) and in SK
(3\%). For the spectrum measurement, we impose further cuts in order to
select \oneRmu events: (c) The event has one ring. (d) The particle
stops inside the 1KT. Among the events after the cuts (a) and (b), 53\%
have one ring. For criteria (d), we require the maximum charge of hit
photomultiplier tubes (PMT) to be less than 200 p.e. This eliminates
effectively the events with muon exiting the 1KT; 68\% of one ring
events retain after this cut. Using this sample, we make the muon
momentum and angular distributions to estimate the spectrum shape of
neutrino, which is described in detail below. The largest systematic
uncertainty for the spectrum measurement in the 1KT is that on energy
scale. We estimate the error on energy scale using cosmic-ray muons and
beam-induced $\pi^0$ sample and quote ${}^{+2}_{-3}$\% for this error.

\subsection{FGD Analysis}
\begin{figure}
 \begin{center}
  \epsfig{figure=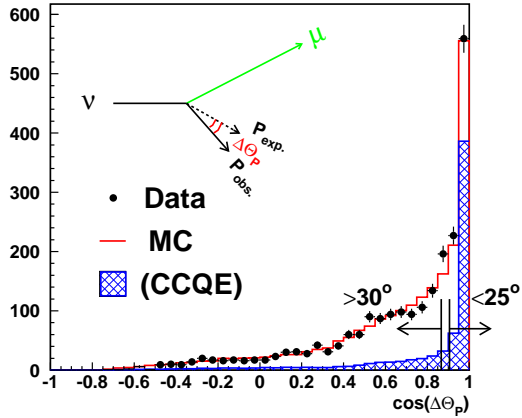,height=60mm}
  \caption{The distribution of $\cos\Dqp$ for 2-track sample in SciFi.
  The schematic explanation for $\Dqp$ is also drawn.
  The events with $\Dqp<25~\mbox{deg.}$ are selected as QE enhanced
  sample, while those with $\Dqp>30~\mbox{deg.}$ are selected as non-QE
  enhanced sample.
  \label{fig:SF-cosdth}}
 \end{center}
\end{figure}
The FGD consist of a scintillating fiber tracker with water target
(SciFi)~\cite{K2K:SciFi-NIM}, plastic scintillator trigger counters
upstream and downstream of SciFi (TRIG), a lead-glass calorimeter (LG),
and a muon range detector (MRD)~\cite{K2K:MRD-NIM}. In the FGD analysis,
we use the events which have one or two tracks with the vertex within
the 5.9~t fiducial volume of SciFi, which is defined as a rectangle of
2.2~m~$\times$~2.2~m in $x$ and $y$, and $1^{\rm st}$ to $17^{\rm th}$
water containers in the beam direction ($z$). The track finding
efficiency is 70\% for a track passing through three layers of SciFi and
close to 100\% for more than five layers~\cite{K2K:SciFi-NIM2}. Three
layers is the minimum track length required in this analysis. Events
which have at least one track passing into MRD are chosen in order to
select \numu-induced CC interactions. Furthermore since SciFi and MRD do
not have fine timing information to select on-spill events, it is
required that the hit in downstream TRIG which matches with SciFi track
should be in spill timing. The momentum of each track is measured by its
range through the SciFi, TRIG, LG and MRD with the accuracy of
2.7\%. The second track also be reconstructed if the proton produced in
the QE interaction has a momentum of typically greater than
600~MeV/$c$. In the case where the second track is visible, the
kinematical information is used to enhance the fraction of QE and non-QE
events in the sample. The direction of second track can be predicted
from the momentum of the first track by assuming QE interaction. The
distribution of cosine of the angular difference between the predicted
and observed second track ($\cos\Dqp$) is shown in
Fig.~\ref{fig:SF-cosdth}. QE enhanced and non-QE enhanced samples are
selected by requiring $\Dqp<25~\mbox{deg.}$ and $\Dqp>30~\mbox{deg.}$,
respectively. The fraction of the QE events in the QE sample is
estimated to be 62\% by MC simulation, while 82\% of events in non-QE
sample is estimated to come from interactions other than QE. The SciFi
events are divided into three event categories; 1-track, 2-track QE
enhanced, and 2-track non-QE enhanced samples. We make muon momentum and
angular distributions for each event category to estimate the neutrino
energy spectrum at ND, which is described below. 

\subsection{Neutrino Spectrum at ND}
\begin{table}[t]
 \begin{center}
  \caption{The central values of the flux re-weighting parameters for the
  spectrum fit at ND (\Flux) and the percentage size of the
  energy dependent systematic errors on \Flux, \FN, and \eSK. The
  re weighting parameters are given relative to the 1.0--1.5 GeV energy bin.
  \label{tab:K2K-fit}}
  \vspace{4mm}
  \begin{tabular}{|c|r|ccc|}\hline
$E_\nu$ (GeV) & \Flux & $\Delta$\Flux & $\Delta$(\FN) & $\Delta$\eSK \\
\hline
0--0.5    &        1.31  & 49   &  2.6 &  8.7 \\
0.5--0.75 &        1.02  & 12   &  4.3 &  4.3 \\
0.75--1.0 &        1.01  &  9.1 &  4.3 &  4.3 \\
1.0--1.5  & $\equiv1.00$ &  --- &  6.5 &  8.9 \\
1.5--2.0  &        0.95  &  7.1 & 10   & 10   \\
2.0--2.5  &        0.96  &  8.4 & 11   &  9.8 \\
2.5--3.0  &        1.18  & 19   & 12   &  9.9 \\
3.0--     &        1.07  & 20   & 12   &  9.9 \\
\hline
  \end{tabular}
 \end{center}
\end{table}
\begin{figure}
 \begin{center}
  \epsfig{figure=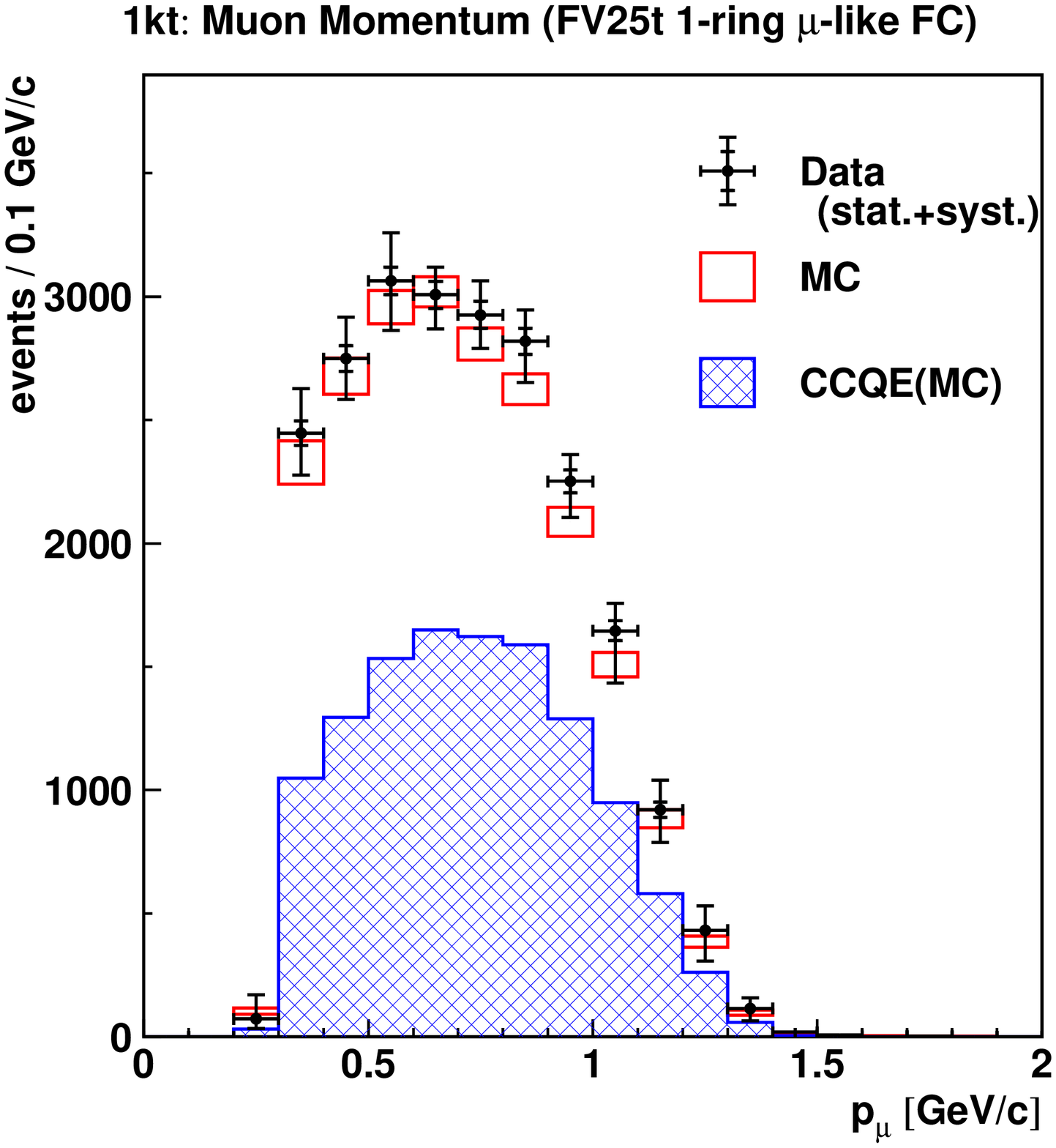,height=65mm}
  \epsfig{figure=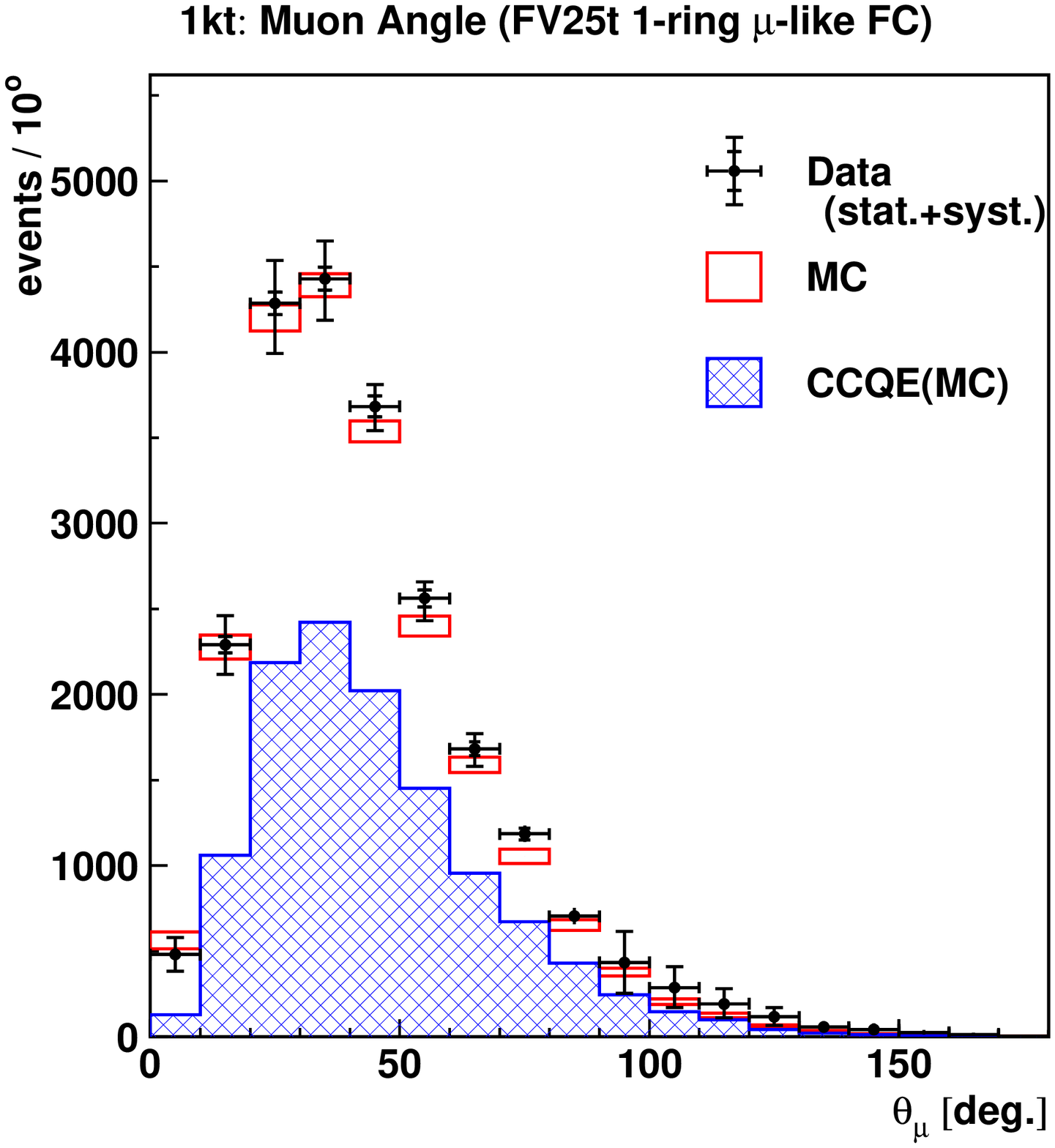,height=65mm}
  \caption{The momentum and angular distributions for 1KT \oneRmu sample.
  The left figure is the distribution of momentum and the right one is
  that of angle with respect to the beam direction. The hatched histograms
  are the contribution from CCQE interaction in MC simulation.
  \label{fig:KT-pmuqmu}}
 \end{center}
 \vspace{15mm}
 \begin{center}
  \epsfig{figure=sfplot-pmu.epsi,height=100mm}
  \hspace{10mm}
  \epsfig{figure=sfplot-qmu.epsi,height=100mm}
  \caption{The momentum and angular distributions for SciFi event samples.
  The left figures are the distributions of momentum and the right ones are
  those of angle with respect to the beam direction for 1-track, 2-track QE
  enhanced, and 2-track non-QE enhanced samples in order of the top, middle,
  and bottom figures. Here again, the hatched histograms are contribution
  from CCQE interaction in MC simulation.
  \label{fig:SF-pmuqmu}}
 \end{center}
\end{figure}
In order to estimate the neutrino energy spectrum at ND, i.e. just after
the production, the 2-dimensional distributions of muon momentum and
angle with the respect to the neutrino beam direction for four event
categories are used: \oneRmu sample in 1KT, 1-track, 2-track QE
enhanced, and 2-track non-QE enhanced samples in FGD. A $\chi^2$-fitting
method is used to compare these data against the MC expectation. The
neutrino spectrum is divided into eight energy bins as defined in
Table~\ref{tab:K2K-fit}. The flux in each energy bin is re-weighted relative
to the value of the beam MC simulation. These re-weighting parameters are
normalized such as the bin of $E_\nu=1.0$--$1.5$~GeV to be 1.0. An
overall normalization is introduced as a free parameter in the fit. The
parameter \Rnqe is used to re-weight the ratio between the QE and non-QE
cross-section relative to the MC simulation. The systematic
uncertainties in the ND measurements, such as the energy scales, the
track finding efficiencies, the detector thresholds, and so on, are
incorporated into the fitting. In addition, PIMON is used in order to
constrain the re-weighting parameters for $E_\nu>1.0$~GeV. 

\begin{figure}
 \begin{center}
  \epsfig{figure=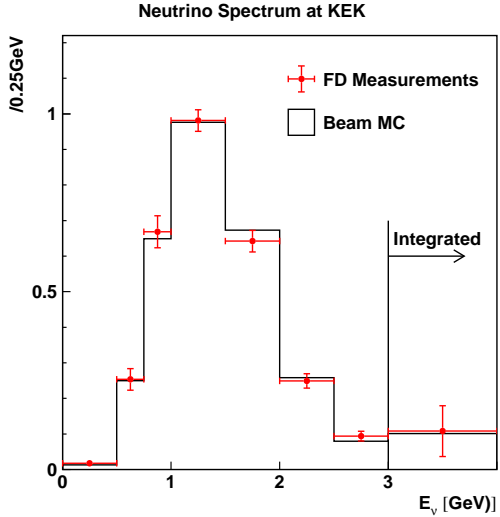,height=70mm}
  \caption{The neutrino energy spectrum at near site inferred from ND
  data along with the result of beam MC simulation. Points with error bars
  are the resulting spectrum of ND fitting, while histogram is the result
  of beam MC simulation.
  \label{fig:ND-enu}}
 \end{center}
\end{figure}

The fitting is successfully done, and the $\chi^2$ value at the best-fit
point is 227.2/197d.o.f. All the parameters including the detector
systematics are found to lie within their expected errors. The best-fit
values of the flux re-weighting factors are shown in
Table~\ref{tab:K2K-fit}. The muon momentum and angular distributions for
\oneRmu sample in 1KT and for 1-track, 2-track QE enhanced and 2-track
non-QE enhanced samples are shown in Fig.~\ref{fig:KT-pmuqmu} and
\ref{fig:SF-pmuqmu} with the re-weighted MC overlaid. The fitting results
are in good agreement with the data. The resulting neutrino spectrum
inferred from the ND data is shown in Fig.~\ref{fig:ND-enu} along with
the beam MC simulation result. As can be seen, the results of beam MC
excellently agrees with ND measurements. The errors on this measurement
are provided in the form of an error matrix, and the correlations between
the parameters are taken into account in the oscillation matrix using
this matrix, later described in detail. The diagonal elements of the
error matrix are shown in Table~\ref{tab:K2K-fit}.

\subsection{Neutrino Interaction Models}
The uncertainty due to neutrino interaction models is studied separately.
In QE interaction, the axial vector mass in the dipole formula is set to
be 1.1~GeV/$c^2$ as a central value and is varied by $\pm10$\%. In single
pion production, the central value of the axial vector mass is set to be
1.2~GeV/$c^2$ and varied by $\pm20$\%~\cite{NEUT:1pi}. For coherent pion
production, two different models are compared: One is the Rein and Sehgal
model~\cite{NEUT:Rein-Sehgal} and  the other is a model by
Marteau \textit{et al.}~\cite{NEUT:Marteau}. For deep inelastic
scattering, GRV94~\cite{NEUT:GRV94} and the corrected structure function
modeled by Bodek and Yang~\cite{NEUT:Bodek-Yang} are studied. Marteau
model for coherent pion production and Bodek and Yang structure function
for deep inelastic scattering are employed in this analysis. The choice
of models causes the fitted value of \Rnqe to change by $\sim$20\%.
Therefore, the error of 20\% on \Rnqe is added to the error matrix. It
is found that the choice of models and axial vector mass does not affect
on the result of spectrum measurement beyond the size of the fitted
errors.

\section{Far-to-near Spectrum Ratio}
The \FN ratio from the beam MC simulation is used to extrapolate the
measurement at ND to those at SK. The detailed description about \FN
can be found in the reference~\cite{K2K:detection}. The errors on \FN
are also provided in the form of an error matrix. The errors including
the correlations on \FN for $E_\nu>1~\mbox{GeV}$ are estimated based on
the PIMON measurements, where PIMON is sensitive in this energy region.
Since PIMON is insensitive in the region of $E_\nu<1~\mbox{GeV}$, the
errors and correlations are estimated based on the uncertainties in the
hadron production models used in K2K beam MC simulation. The errors on
\FN of $E_\nu>1~\mbox{GeV}$ and on that of $E_\nu<1~\mbox{GeV}$ are
treated such that these are not correlated. The diagonal elements in the
error matrix are summarized in Table~\ref{tab:K2K-fit}.

\section{Event Selection in SK}
\begin{figure}
 \begin{center}
  \epsfig{figure=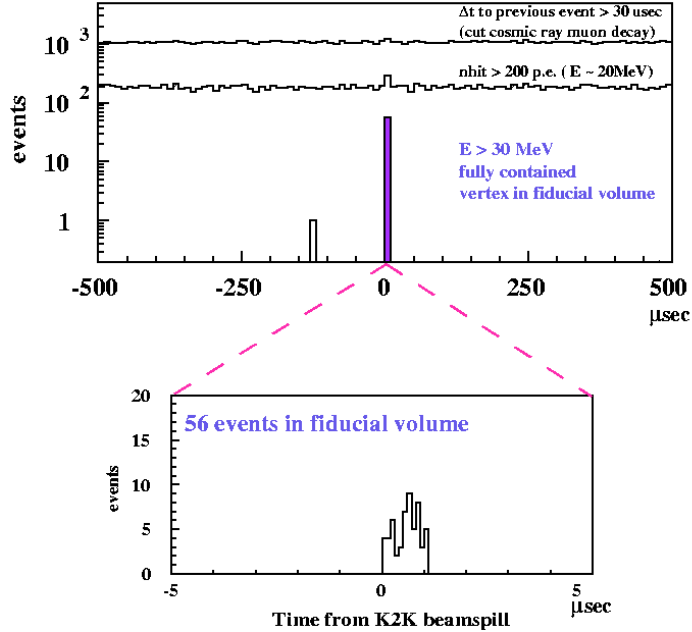,height=85mm}
  \caption{The distribution of $\Delta T (\equiv T_{\rm SK} -
  T_{\rm KEK} - T_{\rm TOF})$ for FC events in SK. The upper figure
  shows the distribution of $\pm500~\mu{\rm sec}$ time window for each
  reduction step in SK. The lower figure show the distribution of
  $\pm5~\mu{\rm sec}$ time window for FCFV events.
  \label{fig:SK-Tdiff}}
 \end{center}
\end{figure}
The selection criteria for SK events are all same as those described in
the reference~\cite{K2K:detection}. Accelerator-produced neutrino
interaction at SK are selected by comparing two Universal Time
Coordinated (UTC) time stamps from the global positioning system (GPS),
$T_{\rm KEK}$ for the KEK-PS beam spill start time and $T_{\rm SK}$ for
the SK trigger time. The time difference between two UTC time stamps,
$\Delta T \equiv T_{\rm SK} - T_{\rm KEK} - T_{\rm TOF}$, where
$T_{\rm TOF}$ is the time of flight of neutrinos from KEK to SK, should
be distributed around the interval from 0 to $1.1~\mu\mbox{sec}$ to
match the width of the beam spill of KEK-PS. The condition
$-0.2<\Delta T<1.3~\mu\mbox{sec}$ is imposed as a timing cut. 
Other criteria are: (a) There is no detector activity  within
$30~\mu\mbox{sec}$ before the event. (b) The total collected p.e. in a
300~nsec time window is greater than 200 ($\sim20~\mbox{MeV}$ deposited
energy). (c) The number of PMTs in the largest hit cluster in the
outer-detector is less than 10. (d) The deposited energy is greater
than 30~MeV. Finally, a fiducial cut is applied to select only the
events with fitted vertices inside 22.5~kt fiducial volume, which volume
is same as that used in SK atmospheric neutrino analysis. The detection
efficiency of this selection is 93\% for charged-current interactions
and 68\% for neutral-current inelastic interactions, for a total of
79\%. Fig.~\ref{fig:SK-Tdiff} shows the $\Delta T$ distribution at the
various stages of the reduction. A clear peak in time with the neutrino
beam from KEK-PS is observed in the analysis time window. After the
above selection, 56 FC events are observed in the fiducial volume
(FCFV events), while the number of FCFV events expected by 1KT
measurement is $80.1^{+6.2}_{-5.4}$ as described in Sec.~\ref{sec:1KT}.
With the timing cut, the expected number of atmospheric neutrino
background is approximately $10^{-3}$ event, which is negligible.

\section{Oscillation Analysis}
A two-flavor oscillation analysis, with \numu disappearance, is performed
with the use of the maximum-likelihood method. In this analysis, both the
number of FCFV events and the energy spectrum shape for \oneRmu are used.
The likelihood is defined as
\begin{equation}
\Ltot = \Lnorm(\Nobs,\Nexp(\dm,\SinSq,f))
\times \Lshape(\dm,\SinSq,f)
\times \Lsys(f).
\end{equation}
The normalization term $\Lnorm$ is the Poisson probability to observe
$\Nobs$ events when the expected number of events is
$\Nexp(\dm,\SinSq,f)$. The spectrum shape term $\Lshape$ is the product
of the probability for each \oneRmu event to be observed at $\Enurec = E_i$:
\begin{equation}
\Lshape = \prod^{N_{1{\rm R}\mu}}_{i=1} P(E_i;\dm,\SinSq,f),
\end{equation}
where P is the normalized $\Enurec$ distribution estimated by MC simulation
and $N_{1{\rm R}\mu}$ is the number of \oneRmu events. The symbol $f$
denotes a set of parameters which is constrained by the systematic errors.
These parameters consist of the re-weighted neutrino spectrum measured by
ND (\Flux), the \FN ratio, the reconstruction efficiency of SK (\eSK) for
\oneRmu events, the re-weighting factor for the QE/non-QE ratio \Rnqe, the
energy scale of SK, and the overall normalization. The errors on first
three items are energy dependent and correlated between each energy bin.
The diagonal elements of their error matrices are summarized in
Table~\ref{tab:K2K-fit}. The error on SK energy scale is 3\%. The
parameters $f$ is treated as fitting parameters with an additional
constraint term $\Lsys$ in likelihood~\footnote{Another approach to treat
the systematic parameters $f$ is performed, which also gives consistent
results with those described in the main text~\cite{K2K:oscillation}.}.

In the oscillation analysis, the whole data sample is used in the
normalization term $\Lnorm$, i.e. $\Nobs=56$. In the spectrum shape term
$\Lshape$, the data taken in June 1999 are discarded since the target
radius and horn current, hence the spectrum shape in June 1999 were
different from the rest of running period. The discarded data correspond
to 6.5\% of total POT. The number of \oneRmu events used in this analysis
is 29, while the number of \oneRmu events from the MC simulation in the
case of no oscillation is 44.
\begin{figure}
 \begin{center}
  \epsfig{figure=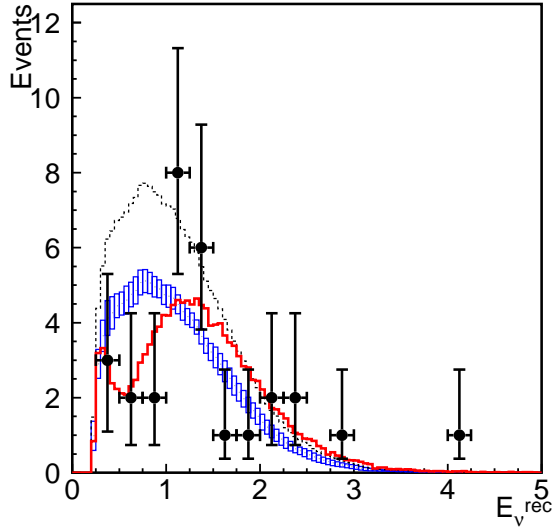,height=70mm}
  \caption{The reconstructed $E_\nu$ distribution for the \oneRmu sample
  (from method 1). Points with error bars are data, while the box
  histogram is the expected spectrum in the case of no oscillation, where
  the height of the box is the systematic error. The solid line is the
  best-fit spectrum in the case where the oscillation occurred. These
  histograms are normalized by the number of observed \oneRmu events (29).
  In addition, the dashed line shows the expectation with no oscillation
  normalized to the expected number of \oneRmu events (44).
  \label{fig:SK-enurec}}
 \end{center}
\end{figure}

The likelihood is calculated at each point in the $\dm$ and $\SinSq$
space to search for the point where the likelihood is maximized. The
best-fit point in the physical region of oscillation parameter space is
found to be at $(\SinSq,\dm)=(1.0,2.8\times10^{-3}~\eV)$. If the whole
space including unphysical region is allowed, these values are found to
be $(\SinSq,\dm)=(1.03,2.8\times10^{-3}~\eV)$. At the best-fit point in
the physical region, the total number of predicted events is 54.2, which
is in agreement with the observation of 56 events within statistical error.
The observed $\Enurec$ distribution for \oneRmu sample is shown in
Fig.~\ref{fig:SK-enurec} together with the expectation for the best-fit
oscillation parameters and that for no oscillation. The
Kolgomorov-Smirnov (KS) test is done to check the consistency between the
observed and best-fit $\Enurec$ spectrum. A KS probability is obtained to
be 79\%, and the best-fit spectrum shape agrees with the observation.

\begin{figure}
 \begin{center}
  \epsfig{figure=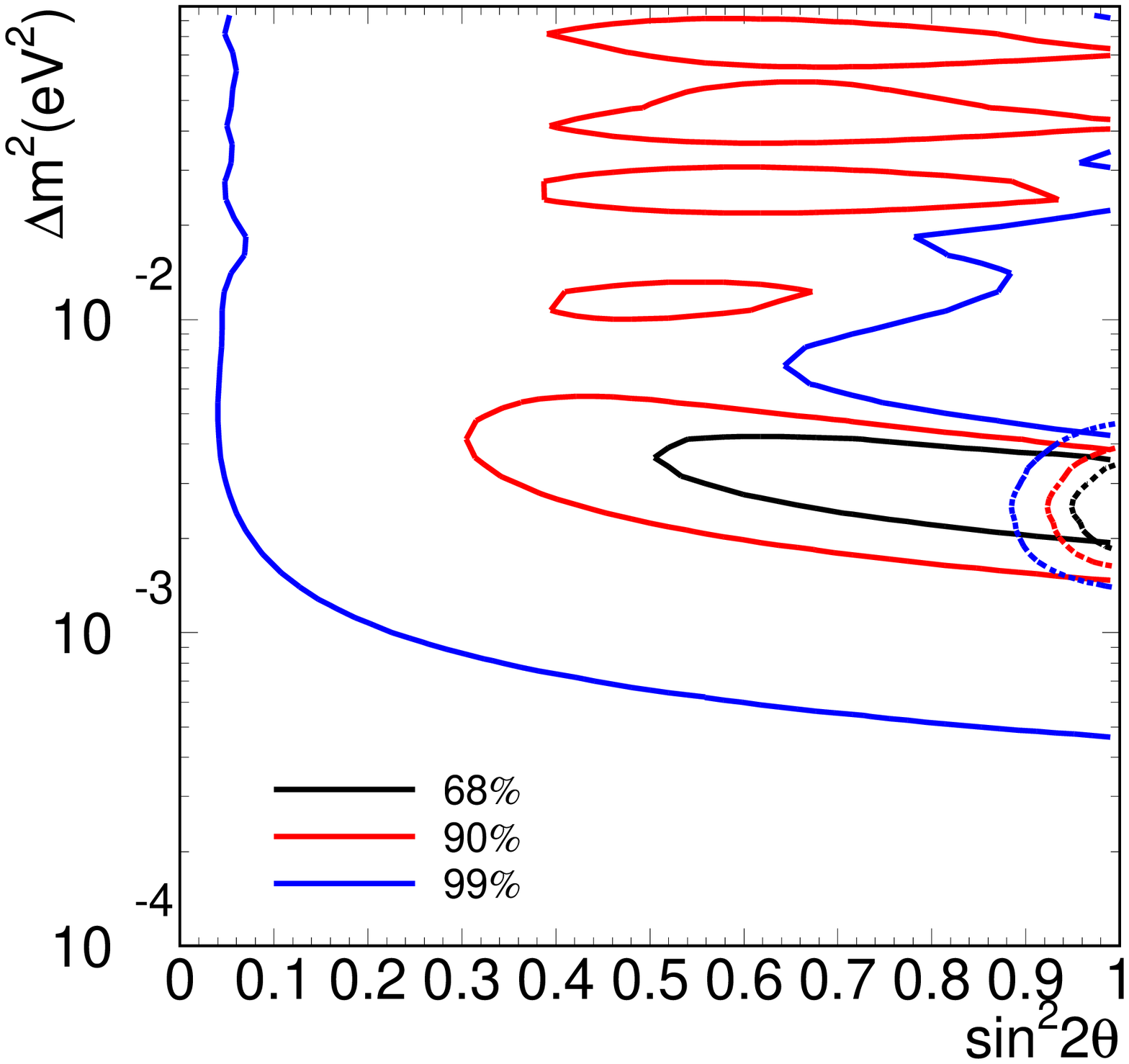,width=0.51\textwidth}
  \hfill
  \epsfig{figure=dL-normshape.epsi,width=0.48\textwidth}
  \caption{Left: Allowed region of oscillation parameters with the
  confidence level of 68\%, 90\%, and 99\%. Solid lines are the results
  from the K2K oscillation analysis. In addition, the results of the
  observation of atmospheric neutrinos by \SK (dashed lines) are
  overlaid. Right: The Behaviors of $-\Delta(\ln\Ltot)$ as a function of
  $\dm$ along $\SinSq=1.0$ axis in the case where normalization-only,
  shape-only, and normalization$+$shape information is used in the
  analysis. The $\dm$ regions indicated by normalization-only and
  shape-only agrees well with each other.
  region as 
  analysis 
  \label{fig:K2K-contour}}
 \end{center}
 \vspace{15mm}
 \begin{center}
  \epsfig{figure=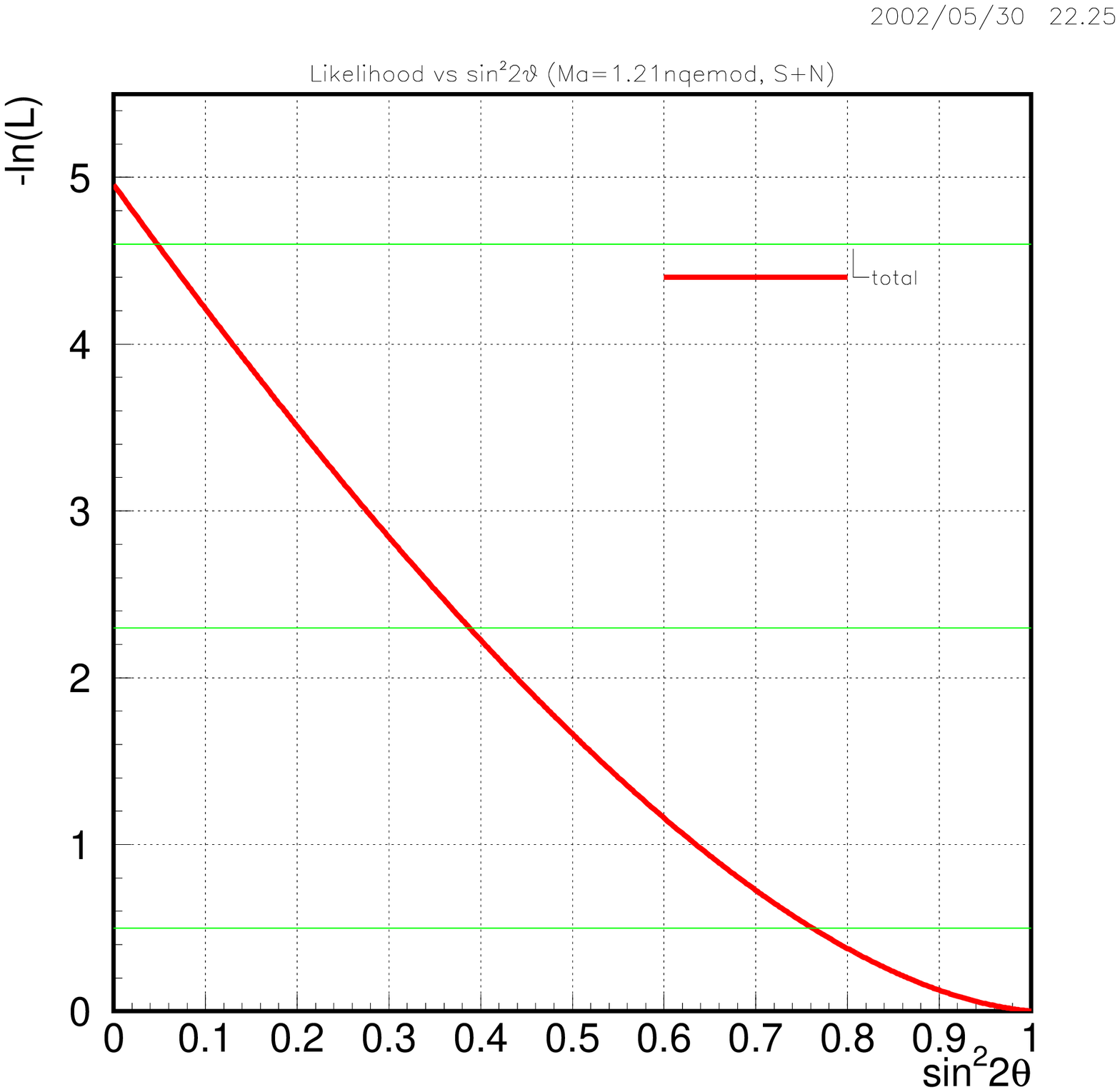,width=0.49\textwidth}
  \hfill
  \epsfig{figure=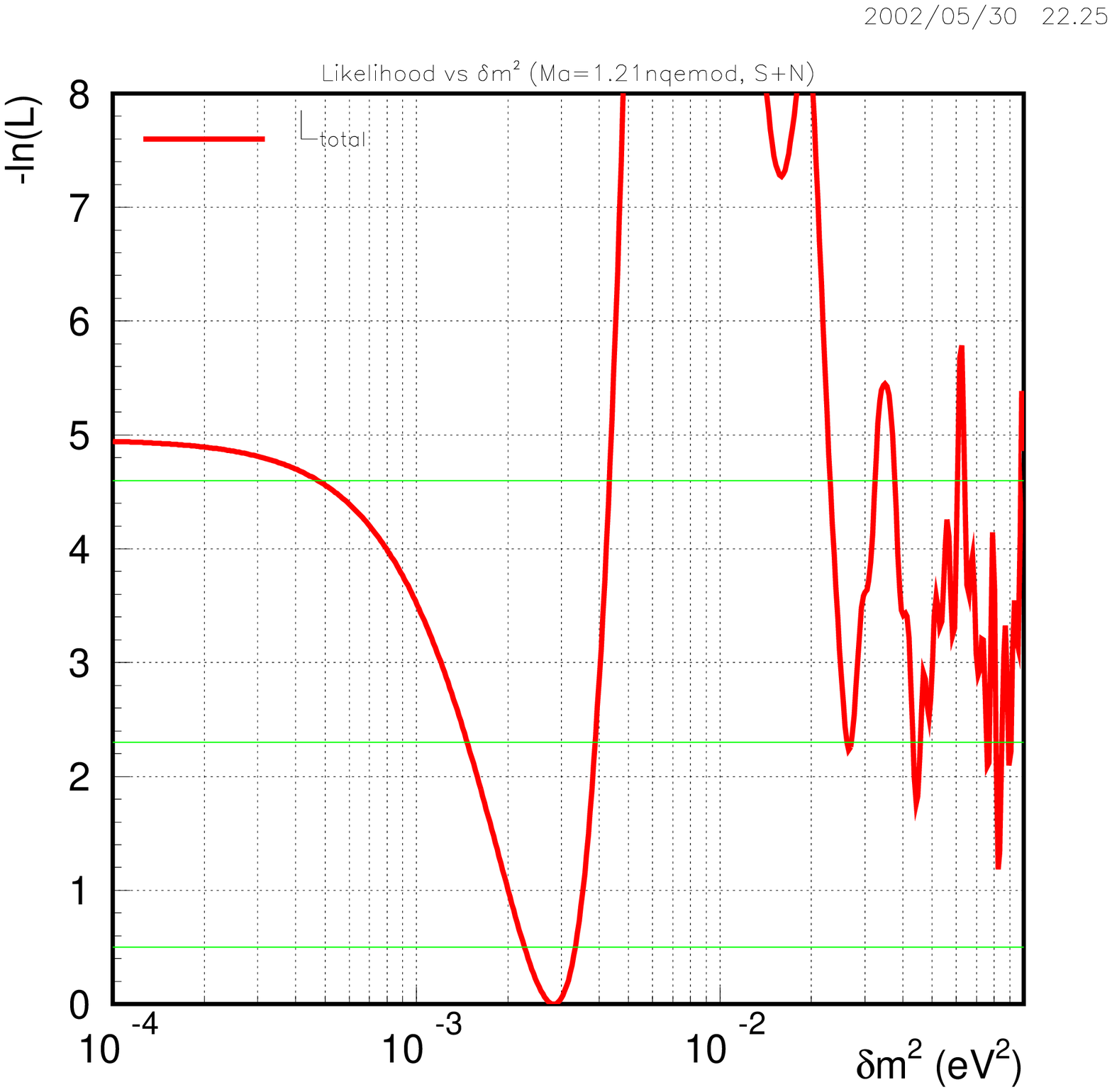,width=0.49\textwidth}
  \caption{The behaviors of $-\Delta(\ln\Ltot)$ as a function of
  $\SinSq$ at $\dm=2.8\times10^{-3}~\eV$ (left) and as a function of
  $\dm$ at $\SinSq=1.0$ (right).
  \label{fig:K2K-dL}}
 \end{center}
\end{figure}

The likelihood ratio of the no oscillation case to the best-fit point is
computed to estimate the probability that the observations are only due
to a statistical fluctuation instead of neutrino oscillation. The no
oscillation probability is calculated to be 0.7\%. When only
normalization or shape information is used, the probabilities are 1.3\%
and 16\%, respectively. The oscillation parameters preferred by the total
flux suppression and the energy distortion alone also agree well with each
other (Fig.~\ref{fig:K2K-contour} (right)). The allowed regions of
oscillation parameters are evaluated by calculating the likelihood ratio
of each point to the best-fit point.

Finally the uncertainties of neutrino interaction models are studied in
the same way as the spectrum measurement at ND. It is found that the
effects of difference between interaction models are negligible for all
the results since the cancellation is caused by using the same models in
ND and SK.

The contours for confidence level of 68\%, 90\%, and 99\% are drawn in
Fig.~\ref{fig:K2K-contour} (left). The behaviors of $-\Delta(\ln\Ltot)$
in the case where normalization-only, shape-only, and
normalization$+$shape information is used in the analysis are plotted in
Fig.~\ref{fig:K2K-contour} (right). This shows that the reduction in the
neutrino flux and the distortion in energy spectrum abserved in K2K
experiment prefer the same $\dm$ region with each other, as described
above. $-\Delta(\ln\Ltot)$ as a function of $\SinSq$ at the best-fit
$\dm$ ($=2.8\times10^{-3}~\eV$) and as a function of $\dm$ at the
best-fit $\SinSq$ (=1.0) are also shown in Fig.~\ref{fig:K2K-dL}. The
90\% C.L. region of $\dm$ is 1.5--3.9$\times10^{-3}~\eV$ at $\SinSq=1.0$.
The allowed region of oscillation parameters obtained by \SK atmospheric
neutrino observations are also drawn in Fig.~\ref{fig:K2K-contour}. The
K2K and \SK results are in good agreement.

\section{The Upgrade of K2K Near Detector --- SciBar Detector ---}
\begin{figure}
 \begin{center}
  \epsfig{figure=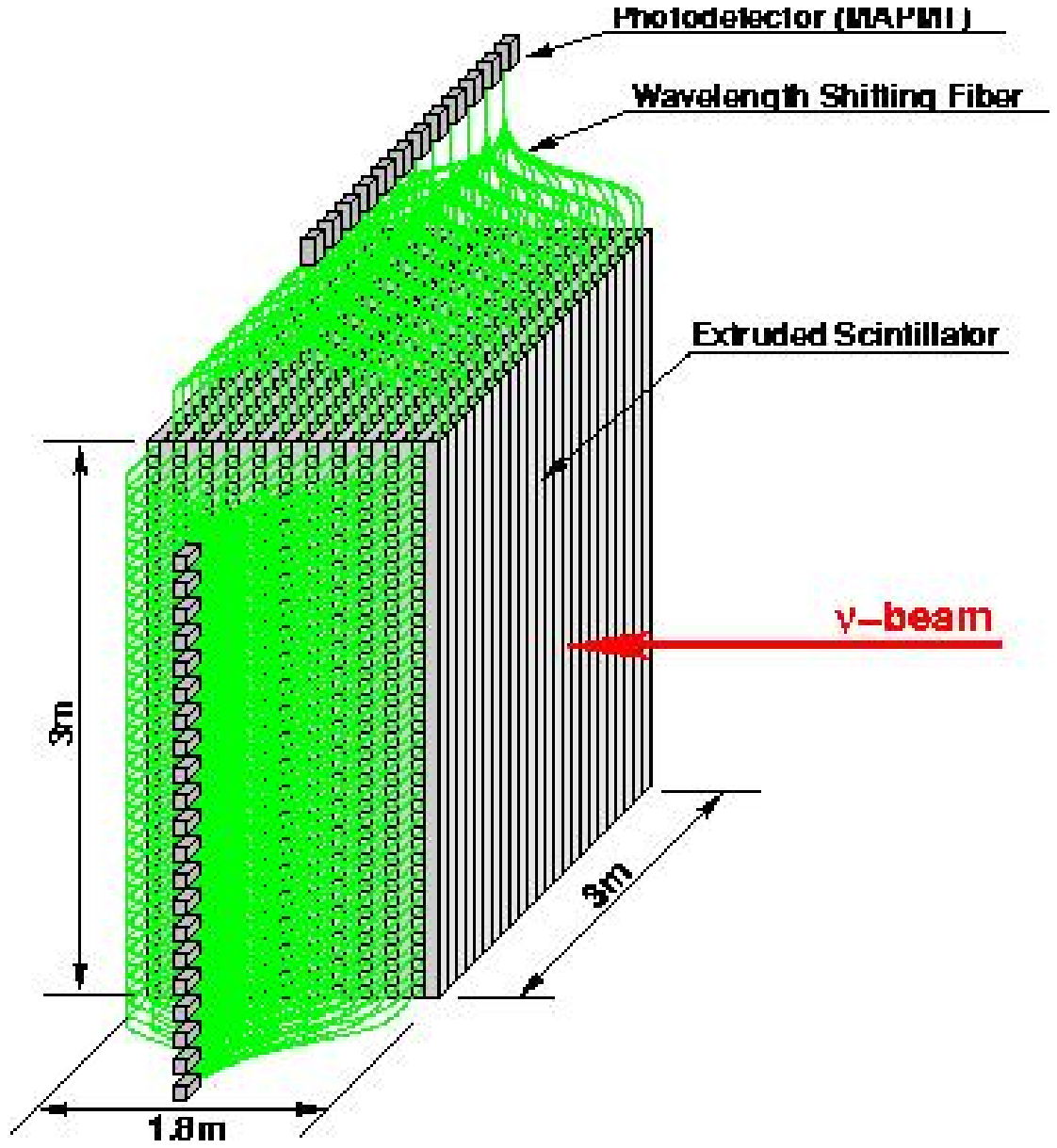,height=90mm}
  \caption{The schematic view of SciBar detector
  \label{fig:K2K-scibar}}
 \end{center}
\end{figure}
In the case where the oscillation parameter is
$\dm\sim3\times10^{-3}~\eV$,
the neutrino energy for oscillation maximum is $\sim0.6~\mbox{GeV}$
under the condition of fixed flight length of 250~km. Therefore, precise
measurement of neutrino spectrum and the knowledge of neutrino
interaction in low energy region are indispensable in order to maximize
the K2K sensitivity. However, low energy particles of which momentum is
below Cherenkov threshold are invisible in the 1KT, while SciFi have
small efficiency for muons below 0.8~GeV/c. 

As an upgrade of K2K near detector, SciBar detector is
designed/developed to detect such low energy particles with high
efficiency. The schematic view of SciBar detector is drawn in
Fig.~\ref{fig:K2K-scibar}. The detector consists of extruded plastic
scintillators which dimensions are 1.3~cm thick, 2.5~cm wide, and 3~m
long. Scintillators are aligned in $X$ or $Y$ to construct a layer
(X-layer or Y-layer). A pair of X- and Y-layer is called as a module,
and 64 modules are aligned in $z$-direction, or beam direction. A hole
of diameter of 1.8~mm is opened at the center of cross section along
through each scintillator. A wavelength shifting fiber is put in this
hole to readout the scintillation light from each scintillator. The
light transmitted outside the scintillator is read by 64ch multi-anode
photomultiplier tubes at one end of fibers. 

The main feature of SciBar detector is that SciBar is a full-active
tracker in which the scintillators play roles of both neutrino
interaction target and particle detector at the same time. Therefore,
SciBar has high efficiency for short tracks, can detect low energy
protons down to 350~MeV/$c$. SciBar also measure $dE/dx$ of each
particle. Using this information, particle identifications and momentum
reconstruction can be performed.

The following study can be done with the use of this excellent detector:
Precise measurement of neutrino spectrum in low energy. Detailed study
of neutrino interactions. Especially non-QE interactions are the
background against neutrino energy reconstruction in CCQE, therefore
to understand these background is very important to maximize the
oscillation sensitivity.

Almost all the R\&D items have been finished and the basic performances
of the detector have been studied. Installation work will be done during
the summer in 2003 and will be finished by the end of September
2003. The data taking will be started from the beam time of October 2003.

\section{Overview of JHF-${\boldmath \nu}$ Experiment}
\begin{figure}
 \begin{center}
  \epsfig{figure=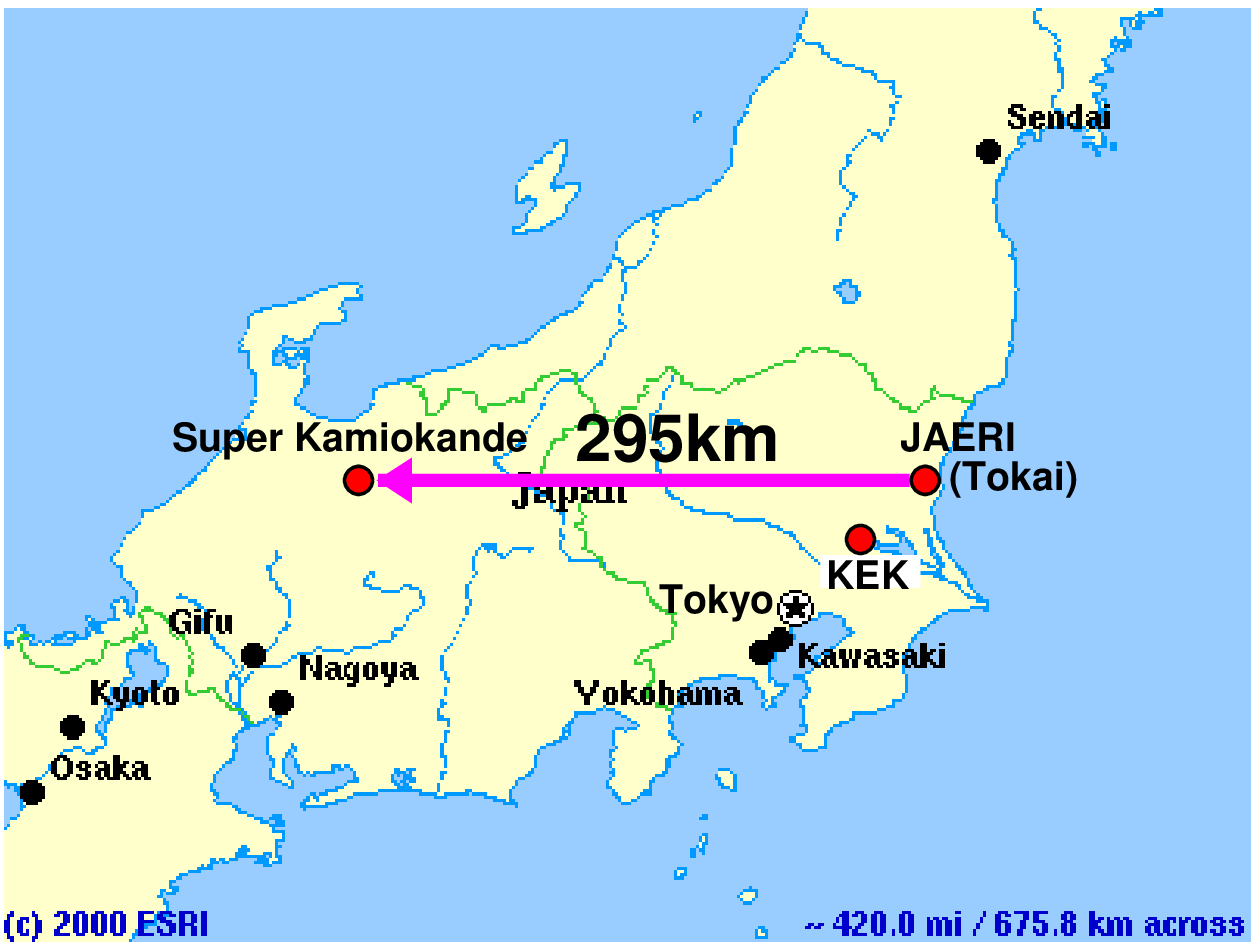,width=0.49\textwidth}
  \hfill
  \epsfig{figure=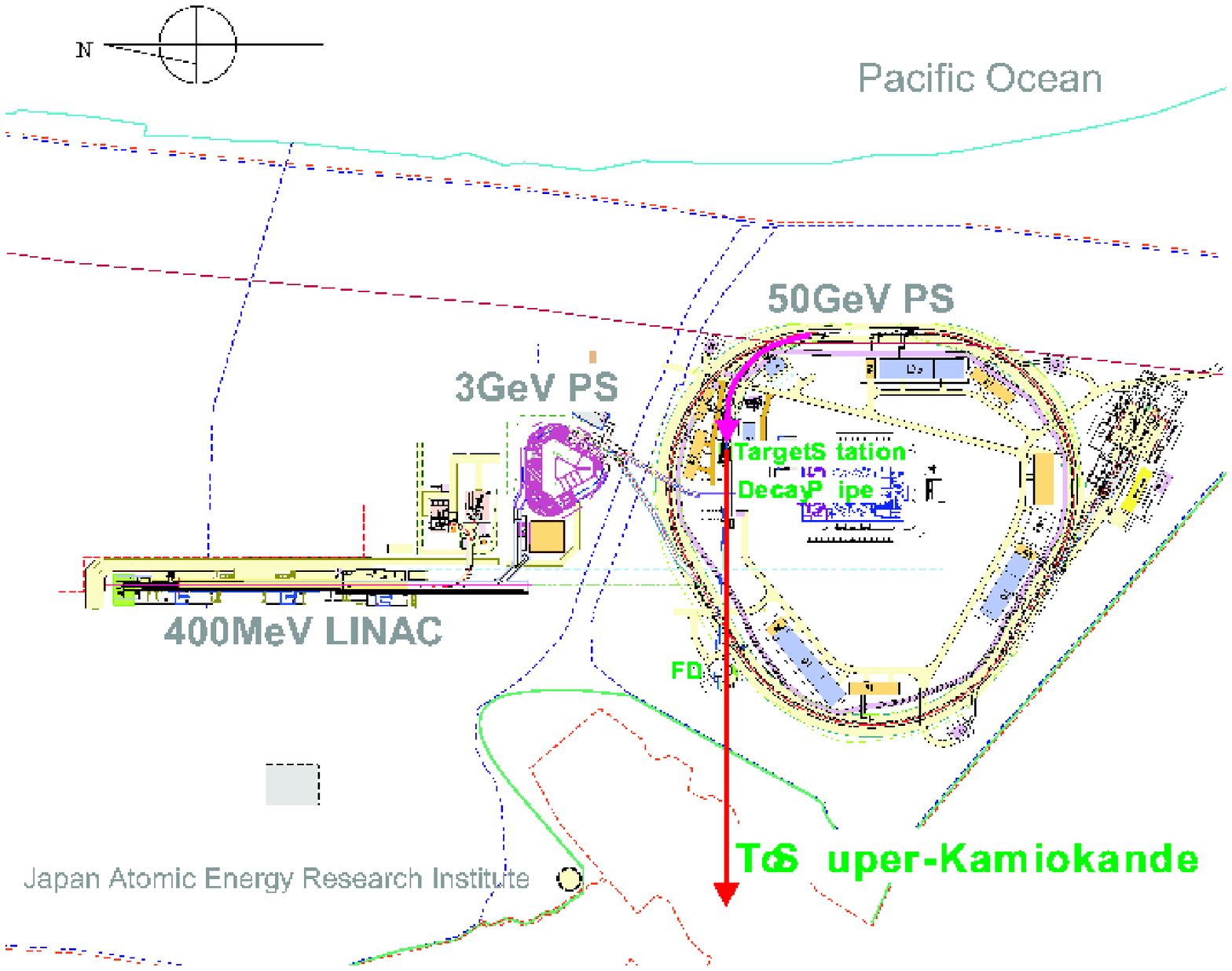,width=0.49\textwidth}
  \caption{The baseline of JHF to Kamioka neutrino project (left) and
  the layout of JHF (right).
  \label{fig:JHF-overview}}
 \end{center}
\end{figure}

The JHF to Kamioka neutrino project is a second generation long
baseline neutrino oscillation experiment (\JHFnu experiment) that
probes physics beyond the Standard Model by high precision
measurements of the neutrino masses and mixing. A high intensity
narrow band neutrino beam is produced  by secondary pions created by
a high intensity proton synchrotron at J-PARC (JAERI). The neutrino
energy is tuned to the oscillation maximum of $\sim$1~GeV for a
baseline length of 295~km toward the world largest water Cherenkov
detector, \SK (left figure of Fig.\ref{fig:JHF-overview}). Its excellent
energy resolution and particle identification enable us to reconstruct
the initial neutrino energy, which is compared with the narrow band
neutrino energy, through the quasi-elastic interaction.

The \JHFnu experiment consists of two phases, phase-I and phase-II.
The physics goal of the phase-I is a search
for the $\nu_\mu\to\nu_e$ appearance with a factor of 20 higher
sensitivity ($\sin^22\theta_{\mu e}\simeq 0.5\sin^22\theta_{13}>0.003$),
the $\nu_\mu\to\nu_\tau$ oscillation measurement with an order
of magnitude better precision ($\delta(\Delta m_{23}^2)=10^{-4}~\eV$
and $\delta(\sin^22\theta_{23})=0.01$) than existing measurements, and a
confirmation of the $\nu_\mu\to\nu_\tau$ oscillation or
discovery of sterile neutrinos (\nus) by detecting the neutral current
events. In the phase-II, an upgrade of the accelerator from 0.75~MW to
4~MW in beam power and the construction of 1~Mt Hyper-Kamiokande
detector at Kamioka site are envisaged. By these upgrade, it is expected
that improvement of another order of magnitude in the
$\nu_{\mu}\to\nu_e$ oscillation sensitivity, a sensitive search
for the CP violation in the lepton sector (CP phase $\delta$ down to
$10^\circ$--$20^\circ$), and improvement of an order of magnitude in the
sensitivity of the nucleon decay search.

The descriptions in this paper are all based on the ``Letter of Intent
for JHF Neutrino Experiment''~\cite{JHF:LOI}. Here only the items related
to neutrino beam of \JHFnu experiment and the search for \nue appearance.
More detailed descriptions are found in the Letter of Intent.

\subsection{Neutrino Beam at JHF}
The layout of JHF is drawn in Fig.~\ref{fig:JHF-overview}(right). The
proton beam is fast-extracted from the 50~GeV PS in a single turn and
transported to the production target. The design intensity of the PS is
$3.3\times10^{14}$ protons per pulse (ppp) at a repetition rate of
0.285~Hz (3.5~sec period). The resulting beam power is 0.75~MW
(2.64~MJ/pulse). The spill width is $\sim5.2~\mu\mbox{sec}$. We define a
typical one year operation as $10^{21}$~POT, which is corresponding to
about 130~days of operation. The protons are extracted toward inside of
the PS ring, and are bent by $90^\circ$ to SK direction in the transport
line using superconducting magnets. The secondary pions (and kaons) from
the target are focused by electromagnetic horns~\cite{JHF:horn}, and
decay in the decay pipe. The length of the decay pipe from the target
position is 130~m. The first near detectors are located at 280~m from
the target. There also exists a plan to put second near detectors
(intermediate detectors) at 2~km from the target.

The beam configuration is off-axis beam (OAB) to produce a narrow neutrino
energy spectrum~\cite{BNL:off-axis}. The optics is almost same as the usual
wide band beam, which is also used in K2K experiment, however, the axis
of the beam optics is displaced by a few degrees from the direction of
far detector (off-axis). A finite decay angle selected, the neutrino
energy becomes almost independent of parent pion momentum as a
consequence of the characteristics of the Lorenz boost, which provides
the narrow spectrum. The peak energy of neutrino can be adjusted by
choosing the off-axis angle. The distance between JHF and SK is
295~km. The neutrino energy will be tuned to between 0.4 and 1.0~GeV,
which corresponds to $\dm$ between 1.6--4$\times10^{-3}~\eV$ suggested
by recent SK and K2K results~\cite{SK:evidence,K2K:oscillation}.
\begin{figure}
 \begin{center}
  \epsfig{figure=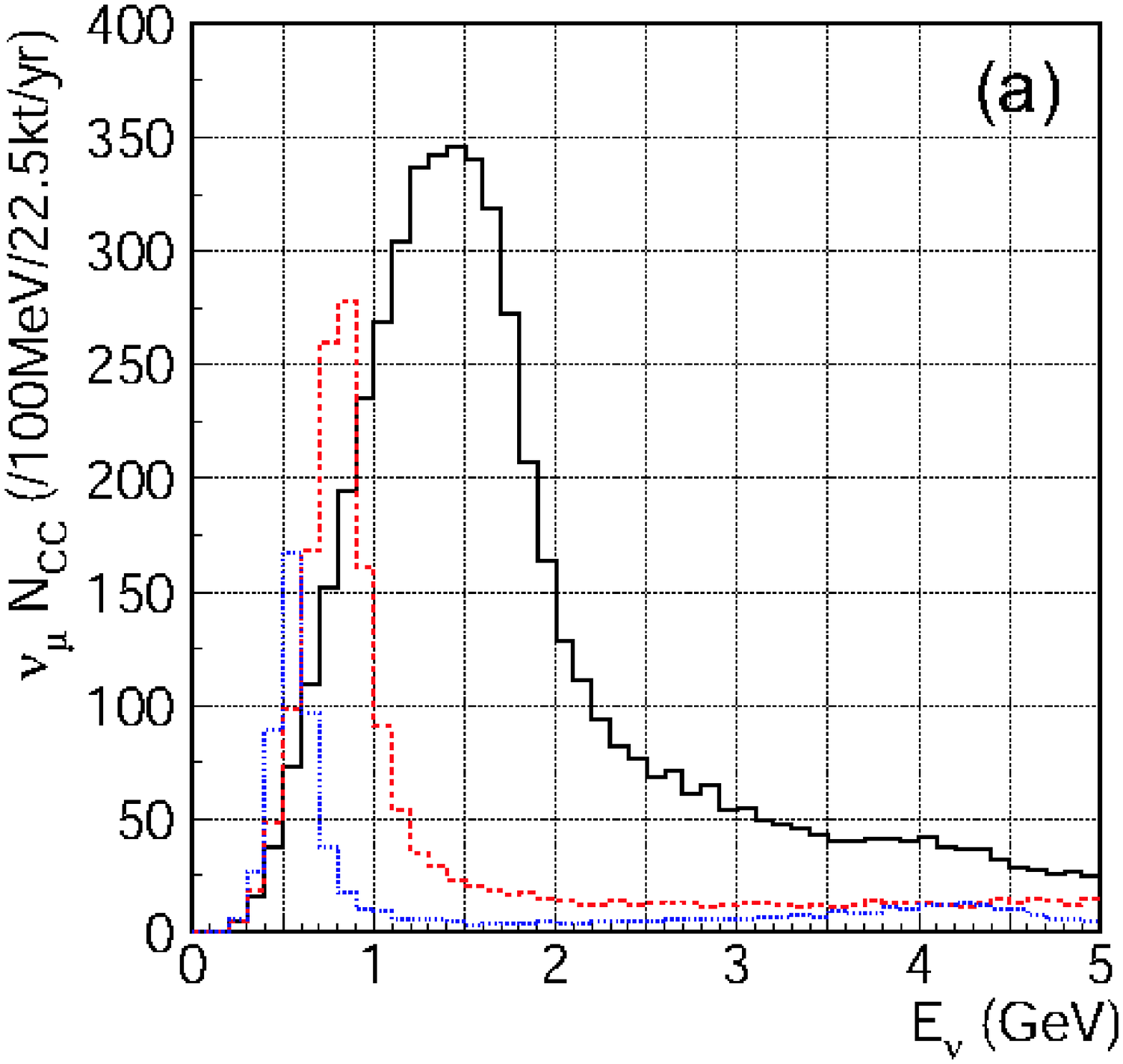,height=70mm}
  \hspace{10mm}
  \epsfig{figure=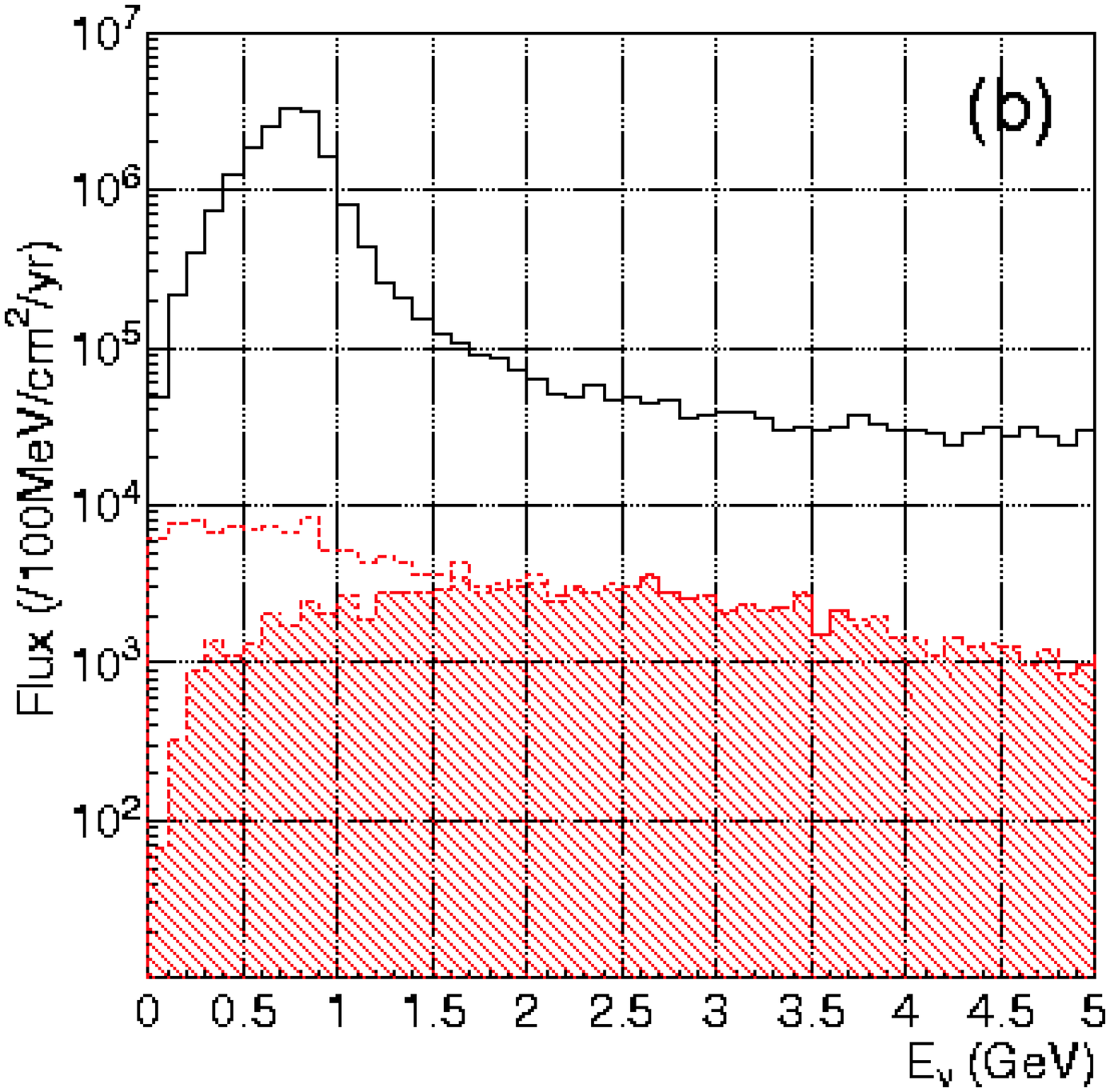,height=70mm}
  \caption{(a) Neutrino energy spectra of charged current
  interactions. Thick solid, dashed and dash-dotted histograms are
  OA1$^\circ$, OA2$^\circ$ and OA3$^\circ$, respectively. (b) Comparison
  of \nue and \numu spectra OA2$^\circ$. Solid (black) histogram
  is \numu and dashed (red) one is \nue. Hatched area is contribution
  from K decay. The low energy $\nu_e$ component is due to $\mu$ decay. 
  \label{fig:JHF-CCspectrum}}
 \end{center}
\end{figure}
Fig.~\ref{fig:JHF-CCspectrum}(a) shows expected neutrino energy spectrum
of charged current interactions at SK in MC simulation. The OAB is
roughly a factor of 3 more intense than usual narrow band beam which is
made by selecting the momentum of pions using dipole magnet placed
between two horns.Fluxes and numbers of interactions are summarized in
Table~\ref{tab:Nint-summary}.

The \nue contamination in the beam is expected to be 1\% at the off-axis
angle of 2$^\circ$ (OA2$^\circ$). The sources of \nue are
$\pi\to\mu\to e$ decay chain and K decay ($K_{e3}$). Their fractions are
$\mu$-decay: 37\%, K-decay: 63\% for OA2$^\circ$. The energy spectra of
the \nue contamination are plotted in
Fig.~\ref{fig:JHF-CCspectrum}(b). At the peak energy of the \numu
spectrum, the \nue/\numu ratio is as small as 0.2\% in OAB. This
indicates that beam \nue background is greatly suppressed (factor
$\sim4$) by applying an energy cut on the reconstructed neutrino energy.
\begin{table}
 \caption{Summary of \numu beam simulation. The peak energy $E_{\rm
 peak}$ is in GeV. The flux is given in $10^6/\mbox{cm}^2/\mbox{yr}$,
 and the \nue/\numu flux ratio is in \%. The ratio in the ``total''
 column is the one integrated over neutrino energy and the column
 ``$E_{\rm peak}$'' is the ratio at the peak energy of \numu
 spectrum. The normalization for the number of interactions are
 /22.5kt/yr. The numbers outside (inside) the bracket are number of
 total (CC) interactions.
 \label{tab:Nint-summary}}
 \begin{center}
  \begin{tabular}{|l|l|rr|rr|rr|}\hline
   & & \multicolumn{2}{c}{Flux} &
   \multicolumn{2}{|c|}{\nue/\numu (\%)}
   & \multicolumn{2}{c|}{\# of interactions}\\
   Beam        & \multicolumn{1}{c}{$E_{\rm peak}$}
   & \multicolumn{1}{|c}{\numu}
   & \multicolumn{1}{c}{\nue}
   & \multicolumn{1}{|c}{total}
   & \multicolumn{1}{c}{$E_{\rm peak}$}
   & \multicolumn{1}{|c}{\numu}
   & \multicolumn{1}{c|}{\nue}\\
   \hline
   OA2$^\circ$ & 0.7 & 19.2 &  0.19 & 1.00 & 0.21 & 3100(2200) & 60(45)  \\
   OA3$^\circ$ & 0.55 & 10.6 & 0.13 & 1.21 & 0.20 & 1100(~800) & 29(22)  \\
   \hline
  \end{tabular}
 \end{center}
\end{table}

\subsection{Search for \nue Appearance}
The JHF neutrino beam has small \nue contamination (0.2\% at the peak
energy of OAB) and the $\nu_e$ appearance signal is enhanced by tuning
the neutrino energy to be its expected oscillation maximum. Therefore,
\JHFnu experiment has an excellent opportunity to discover \nue
appearance and hence to measure $\theta_{13}$. The sensitivity on \nue
appearance is studied based on the full MC simulations and analysis of
SK and K2K experiments. 

The \nue appearance signal is searched for in the CCQE interaction, for
which the energy of neutrino can be calculated using kinematics. Since
the proton momentum from the QE interaction is usually below the
Cherenkov threshold, the signal has only a single electro-magnetic
shower ring (\oneRe).

\begin{figure}
 \begin{center}
  \epsfig{file=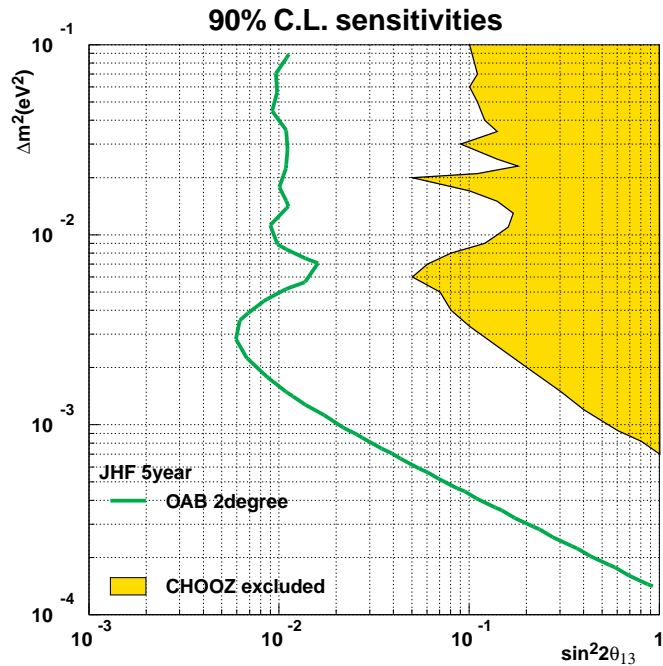,width=95mm}
  \caption{The 90\% C.L. sensitivity contours in 5 years of operation
  with OA2$^\circ$. The 90\% C.L. excluded region of CHOOZ is plotted as
  a comparison. $sin^2\theta_{23}$ is assumed to be 0.5 and the possible
  contribution due to $\theta_{12}$ term is assumed to be small compared
  to the one due to $\theta_{13}$ term. 
\label{fig:JHF-contours}}
 \end{center}
\end{figure}

The standard SK atmospheric neutrino analysis criteria are used to
select \oneRe events: single ring, electron like (showering), visible
energy greater than 100~MeV, and no decay electrons. Reduction of number
of events by the ``standard'' 1-ring $e$-like cut for charged and neutral
current events are listed in Table~\ref{tab:cuteff_oa2}. The excellent
$e/\mu$ separation capability and $\mu\to e$ detecting capability are
key features of the effective elimination of \numu charged current and
all of the inelastic events which contain charged $\pi$. The remaining
background events at this stage are predominantly from single $\pi^0$
production through neutral current interactions and from \nue
contamination in the beam.
\begin{table}
 \begin{center}
  \caption{Number of events and reduction efficiency of ``standard''
  1-ring e-like cut and $\pi^0$ cut for 5 year exposure ($5 \times
  10^{21}$ POT) OA$2^\circ$. For the calculation of oscillated
  \nue, $\dm=3\times10^{-3}~\eV$ and $\SinSq_{\mu e}=0.05$ is assumed.
  \label{tab:cuteff_oa2}}
  \begin{tabular}{|l|rrrr|}\hline
   OAB $2^\circ$ & \numu CC & \numu NC & Beam \nue & Oscillated \nue \\ 
   \hline 
   (1) Generated in F.V. & 10713.6 & 4080.3 & 292.1 & 301.6 \\
   (2) 1R e-like & 14.3 & 247.1 & 68.4 & 203.7 \\
   (3) e/$\pi^0$ separation & 3.5 & 23.0 & 21.9 & 152.2 \\
   (4) $0.4~\mbox{GeV}<\Enurec<1.2$~\mbox{GeV} & 1.8 & 9.3 & 11.1 & 123.2  \\
   \hline
  \end{tabular}
 \end{center}
\end{table}
In order to reduce the background events which come from $\pi^0$,
further cuts are applied based on following information; (1) angle
between $\nu$ beam direction and $e$-like ring, (2) invariant mass of two
photons assuming that the event contains two rings which cannot be
separated by ``standard'' algorithm, (3) difference between single and
double ring likelihood, (4) energy fraction of lower energy ring,
$\frac{E(\gamma_2)}{E(\gamma_1)+E(\gamma_2)}$, assuming the event
contains two rings. Table~\ref{tab:cuteff_oa2} lists the number of
events after this $e/\pi^0$ separation.
Extra rejection of an order of magnitude (23/247.1) in the \numu neutral
current background is achieved with the signal acceptance of
152.2/203.7=75\%. 
\begin{figure}
 \begin{center}
  \epsfig{file=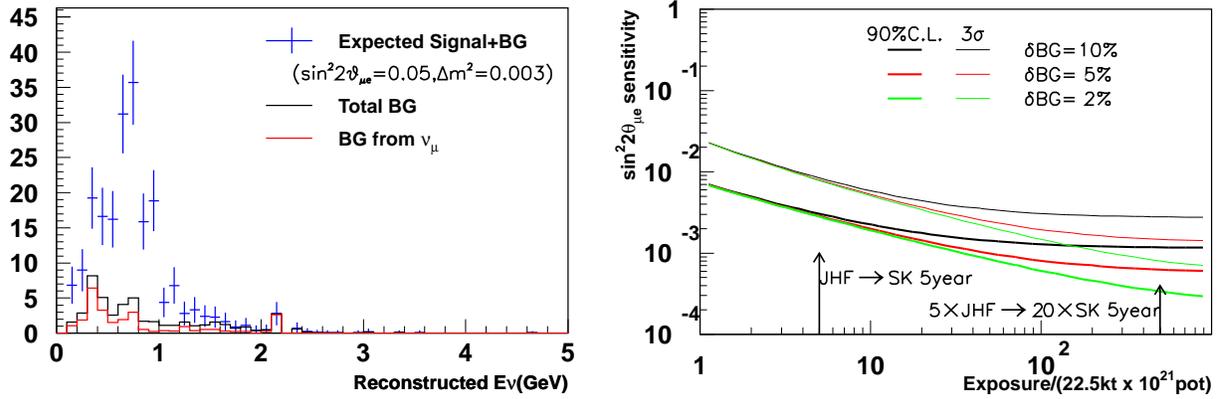,width=\textwidth}
  \caption{(Left) Expected reconstructed neutrino energy distributions
  of expected signal+BG, total BG, and BG from \numu interactions for 5
  years exposure of OA2$^\circ$. (Right) Expected 90\%CL sensitivity
  (thick lines) and $3\sigma$ discovery contours (thin lines) as the
  functions of exposure time of OA2$^\circ$. In left figure, expected
  oscillation signals are calculated with the oscillation parameters of
  $\dm = 3\times10^{-3}~\eV$ and $\sin^22\theta_{\mu e}$ (a effective mixing
  angle $ = \sin^2\theta_{23}\cdot\sin^22\theta_{13}$) $=0.05$. In the
  right figures, three different contours correspond to 10\%, 5\%, and
  2\% uncertainty in the background estimation.
  \label{fig:sens_oa2.0}}
  \end{center}
\end{figure}
Fig.~\ref{fig:sens_oa2.0} (left)
shows the reconstructed neutrino energy distributions for 5~years. The
oscillation parameters of $\dm = 3\times10^{-3}~\eV$ and
$\sin^22\theta_{13}=0.1$ are assumed. A clear appearance peak is seen at
the oscillation maximum of $E_\nu\sim0.75~\mbox{GeV}$.
The right plot of Fig.~\ref{fig:sens_oa2.0} show 90\% and 3$\sigma$
limits as a function of the years of operation with the systematic
uncertainty of background subtraction to be 2\%, 5\% and 10\%. The
sensitivity of $\sin^22\theta_{13}=0.006$ at 90\% confidence level can
be achieved in five years of operation. Fig.~\ref{fig:JHF-contours}
shows 90\%C.L. contours of sensitivity for 5 year exposure assuming 10\%
systematic uncertainty in background subtraction.

\section{Summary}
In the K2K experiment, both the number of observed neutrino events and
the observed energy spectrum at SK are consistent with neutrino
oscillation. It is concluded that the probability that our measurements
are explained by statistical fluctuation without oscillation is less
than 1\%. The best-fit point in the oscillation parameter space is
$(\SinSq,\dm)=(1.0,2.8~\eV)$ and the 90\% C.L. region of $\dm$ at
$\SinSq=1.0$ is 1.5--$3.9\times10^{-3}~\eV$, which agrees well with the
atmospheric neutrino observations by \SK. After the accident of \SK in
November 2001, the reconstruction work have been done by the beginning
of December 2002, and K2K is now running, as K2K-II, to collect 2 times
more data, and finally up to planned $\sim10^{20}$ POT. As the upgrade
of K2K near detectors, SciBar detector will fully installed in summer of
2003. The SciBar detector will enable us to study lower energy neutrino
interactions and cross-section of neutrino interactions.

In order to confirm precisely the results of the first generation
neutrino experiments and to explore the physics beyond the Standard
Model, JHF to Kamioka neutrino experiment, \JHFnu experiment, is
planned. With the use of high intensity neutrino beam, $\nu_\mu\to\nu_e$
oscillation will be explored down to $\sin^22\theta_{13}=0.006$ at 
$\dm\sim3\times10^{-3}~\eV$ with 90\% C.L., which is 20 times higher
sensitivity than CHOOZ experiment. The R\&D has already started and the
experiment is planned to start in 2007.

\section*{Acknowledgments}
I gratefully acknowledge Prof. Y. Totsuka for his financial support to
give me a precious opportunity to attend such a authoritative conference
and give a talk there. I also thank the Japan Society for Promotion of
Science (JSPS) for the support. The K2K experiment have been build and
operated with the support from the Ministry of Education, Culture,
Sports, Science and Technology, Government of Japan, by the
U.S. Department of Energy, by the Korea Research Foundation, and by the
Korea Science and Engineering Foundation.


\end{document}